\documentclass[ 
 reprint,
 amsmath,amssymb,
 aps,
]{revtex4-2}

\usepackage{graphicx}
\usepackage{dcolumn}
\usepackage{bm}
\usepackage{subcaption}
\usepackage{ragged2e}
\usepackage{hyperref}

\hypersetup{
           breaklinks=true,   
           colorlinks=true,
           urlcolor=[RGB]{0,0.50,1.0},
           linkcolor=[RGB]{0,0.50,1.0}
            }

\begin{document}

\preprint{prl}

\title{Experimental Signatures for Identifying Distinct Origins of Color Field Generation}

\author{Subramanya Bhat K. N.}
\altaffiliation[]{}
\email{subramanyabhatkn@gmail.com}
\author{Amita Das}%
\email{amita@iitd.ac.in}
\affiliation{%
 Indian Institute of Technology Delhi, New Delhi, India 
}%
\author{Bhooshan Paradkar}
\affiliation{%
  UM-DAE Centre for Excellence in Basic Sciences, University of Mumbai 
}%
\author{V Ravishankar}
\email{vravi@iitd.ac.in,v.ravishankar@iiitdwd.ac.in}
\affiliation{Indian Institute of Information Technology Dharwad, Dharwad, India}%



\begin{abstract}
Signatures for non-abelian dynamics have long been central to QCD and QGP.  Equally important are they in spin systems and laser-plasma interactions where they emerge as effective interactions.  Distinguishing experimentally gauge inequivalent sources (and hence potentials) that produce the same field tensor is one major task in this endeavour.  As a step in this direction, this paper investigates how physically distinct sources which produce the same color electric field (uniform and constant) may be distinguished experimentally in a gauge-invariant manner. We first study the motion of a test particle in such fields and show that the resultant trajectories are counterintuitive. We then examine the radiation emitted —  both gluonic and photonic and show that each source (independent non-abelian configuration)  leaves a unique signature in the energy spectra, laying the ground for application to specific physical systems. 

\end{abstract}

\maketitle

Yang-Mills (YM) theories, as non-abelian gauge theories are known, are ubiquitous in physics. Introduced initially to understand the dynamics of strong interactions \cite{YangMills-original} (and even earlier, to describe a Dirac particle in a gravitational field \cite{SchUnification}), they have found applications in diverse phenomena such as spintronics and interaction of cold atoms with laser fields \cite{Leurs2007,ColdAtomPRLOsterloh,ColloquiumArtificialG,Berche_2013}. Generation of synthetic non-abelian gauge fields is now commonplace and their topological effects are quite extensively studied both experimentally and theoretically. These studies are largely “classical” in the sense that the fields are not quantised. Furthermore, the physics of non-abelian fluids is important to understand quark matter in neutron stars and the evolution of quark-gluon plasma in heavy ion collisions. Pertinently, it has been argued that in the long wavelength quasi-static regime, a classical description of gauge theories emerges naturally \cite{HUET199794, BUCHMULLER1998219}. Thus, for example, it was recently shown that a nonabelian plasma can exhibit novel instabilities \cite{KNmine}. In short, YM theories merit equal attention in classical and quantum regimes.

In reality, even elementary features of classical YM (CYM) theories leave a lot to be explored. Though efforts have been put into obtaining non-abelian solutions to YM equations, for example in \cite{ScreenSolSikivieWeissPRL,YMSikivieWeiss,StatSourceSikivieWeiss,PotfromEBWeiss, ActorSols, FROLDI2022169026}  its counterpart — test particle dynamics in external fields, has received relatively scant attention. At the heart of this exercise lie two defining features of YM dynamics: (i) Unlike as in electrodynamics, the field tensor does not furnish complete information on the system \cite{WuYang}; physically distinct sources may produce the same field tensor.  (ii) The test charge (designated as color) is a dynamical variable and obeys Wong equation which involves gauge potentials \cite{WongClassicalYM-Isospin}. Finally, in the case of strong interactions, there is a further restriction that only gauge invariant quantities may be measured experimentally. Thus, the question of whether/ how test particle motion can unravel information on the underlying sources requires a deeper look.  This letter addresses this question with a seemingly simple but illustrative example of charged particle dynamics in a non-abelian static color electric field. The field is so chosen that it may be produced by an infinite one-parameter family of potentials, all of which are gauge inequivalent to each other. Some relevant features of this problem have been discussed in \cite{LSBrown}, albeit qualitatively.  In answer to the question posed, we show that the nature of the potential (and hence the source)  clearly gets imprinted in particle dynamics. We further show that the accompanying bremsstrahlung radiation (be it photonic or gluonic)  also carries the signature for each source in its power spectrum. We expect that our results, when combined with a similar exercise with color magnetic field, will be of use in gaining a better appreciation of YM dynamics from spintronics and atom laser interactions to the  QGP produced in heavy ion collisions.

As mentioned, we consider test particle dynamics in a non-abelian static color electric field. To keep the analysis simple, we shall take the internal color space to be $\mathbb{R}^3$ and the gauge (color) group to be $SO(3)$, the set of all rotations in $\mathbb{R}^3$. This choice is not unrealistic since such fields do arise in spin systems \cite{Berche_2013}. Since the color charge and the color fields transform as vectors under gauge transformations, we follow \cite{KNmine} and employ for them the notations $ \vec{q},~ \vec{E}_i$ where the subscript refers to spatial components and the vector sign to the color degrees of freedom. Though non-covariant, the underlying scalar and vector potentials are also represented, for conciseness,  as  $\{ \vec{\phi}, \vec{A}_i\}$.  In what follows, $\{\hat{e}_1,\hat{e}_2,\hat{e}_3\}$ represents an orthonormal basis in the color space.

Let us begin with the basic relation
\begin{equation}
    \label{feq}
    \vec{E}_i   =   -\partial_{i}\vec{\phi} -\frac{1}{c}\partial_0 \vec{A}_{i} - g \vec{\phi} \times \vec{A}_{i} .
\end{equation}
The term multiplied by the nonabelian coupling constant $g$ is the hallmark of YM  dynamics.  Hence,  we consider the class of color electric fields which, in a suitable gauge, may be generated entirely by the cross-product term in Eq. (\ref{feq}) with constant $ \{ \vec{\phi}, \vec{A}_i\} $.  The resultant fields are thus maximally non-abelian. Our objective  is accomplished by  the class of potentials characterised by a real parameter $\alpha > 0$ :
\begin{equation}
    \label{gpc}
    \vec{\phi}^{\alpha} = \alpha \phi \hat{e}_3;~ \vec{A}_x^{\alpha} =\frac{1}{\alpha}  A\hat{e}_2;~ \vec{A}_y = \vec{A}_z =0
\end{equation}
which yields 
\begin{equation}
\label{eform}
\vec{E}_x = g\phi A \hat{e}_1;~ \vec{E}_y = \vec{E}_z =0; \vec{B}_i \equiv 0
\end{equation}
Remarkably, the resulting color electric field is independent of $\alpha$, thereby exhibiting the Wu-Yang ambiguity \cite{WuYang} vividly. In short, the set $ \{ \vec{\phi}^{\alpha}, \vec{A}_i^{\alpha}\} $ constitutes a one-parameter family of mutually gauge inequivalent potentials all of which generate the same field. None of them can be generated by a Maxwellian source.

The dynamics of the test particle (mass $m$, color charge $\vec{q}$) is governed  by the Lorentz and the Wong equations (we drop the spatial labels since the motion is one-dimensional),
\begin{equation}
    \label{eqm}
    \frac{d P}{dt}   =   \vec{Q}\cdot  \vec{E};~  \frac{d\vec{Q}}{dt} = g c \vec{Q} \times \left( \vec{\phi} - {\vec{A}}\frac{v}{c}\right) 
\end{equation}
in writing which we have employed $\vec{Q} \equiv g \vec{q}$ which is the exact analogue of the electric charge. In fact, the abelian limit ensues when $g \rightarrow 0 $, $q \rightarrow \infty$ with $Q = gq$ held constant. Eq (\ref{eqm}) may be treated either as a classical equation or as that of the corresponding operators in the Heisenberg representation. In this first study, we shall treat it as completely classical, after noting that the adaptation to the quantum mechanical situation is quite straightforward. 

To facilitate further study we shall express the dynamical quantities in units of the scales furnished by the system. First and foremost is the dimensionless quantity
\begin{equation}
    k = \frac{gq^2E}{\left(mc^2\right)^2}
\end{equation}
which plays a dual role. It acts as the effective coupling constant that characterises the nonabelian dynamics, and is also a measure of the unique system energy scale $q \sqrt{gE}$ relative to the rest mass energy of the particle. Together  with other   intrinsic scales, we set up  the dimensionless variables, defined by
\begin{eqnarray}
    t &  =  & \left(\frac{mc}{Q E}\right) \tau;~  v = uc; ~ {\cal E}  = \left(mc^2 \right)\varepsilon; P = \left(mc\right)p\nonumber \\
     \phi & =  & \left(\frac{kmc^2}{Q}\right)\varphi; ~A = \left(\frac{kmc^2}{Q}\right) a ;~ \vec{\zeta} =  \frac{\vec{Q}}{Q},
\end{eqnarray}
in terms of which, Eq (\ref{eqm}) gets recast as
\begin{eqnarray}
    \label{eoms}
     \frac{dp}{d\tau} & = & \zeta_1;~
     \frac{d\zeta_1}{d\tau}  =    (\zeta_2\varphi + \zeta_3au) \nonumber  \\
     \frac{d\zeta_2}{d\tau} & = &  - \zeta_1\varphi: ~ 
    \frac{d\zeta_3}{d\tau} = -\zeta_1au
\end{eqnarray}
The set of equations (\ref{eoms}) is integrable. The orbit of the test charge is characterised by three constants of motion which, apart from $ Q $, are given by the total energy and the canonical momentum:
\begin{equation}
    \label{eq:consts}
       {\cal \varepsilon}  =  \sqrt{1+ p^2} + \zeta_3 \phi;  ~~
        \Pi  =  p + \zeta_2 a
\end{equation}
employing which, we arrive at the defining equation 
\begin{equation}
    \label{effdyn}
    \frac{d^2 p}{d\tau^2}  + k_1 p  - k_2 \frac{p}{\sqrt{1+p^2}} - k_3 =0; 
\end{equation}
where the constant coefficients are given by
\begin{equation}
    \label{eq:k_consts}
    k_1 = \left(\alpha^2+ \frac{1}{\alpha^2}\right);~ 
    k_2 = \frac{1}{\alpha^2} \varepsilon; ~ 
    k_3 = \alpha^2 \Pi
\end{equation}

\begin{figure}[h!]
    \centering
    \includegraphics[width=0.9\linewidth]{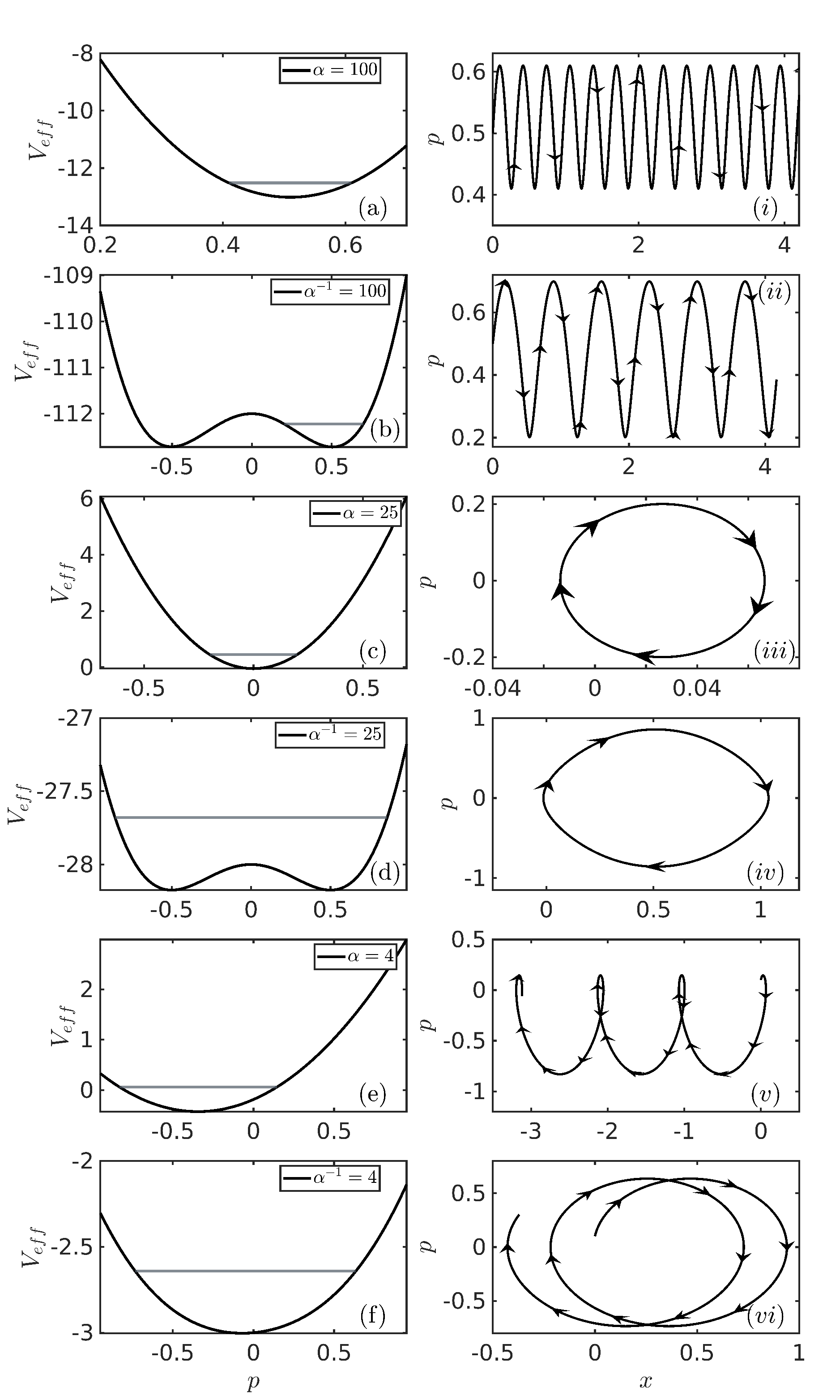}
    \caption{\justifying{Plots showing phase space plots (right) corresponding to different $V_{eff}(p)$ (left). The parameters are, for the top two figures  ($ \alpha  (\alpha^{-1}) =100 , \varepsilon = 1.12, ~ \Pi = 0.5$), for the middle two figures,  ($ \alpha  (\alpha^{-1}) =25 , \varepsilon = 1.12, ~ \Pi = 0.0$), and for the bottom two, ($ \alpha  (\alpha^{-1}) = 4 , \varepsilon = 0.75, ~ \Pi = -0.35$) }}
    \label{fig:1}
\end{figure}

\begin{figure}[h!]
    \centering
    \includegraphics[width=0.8\linewidth]{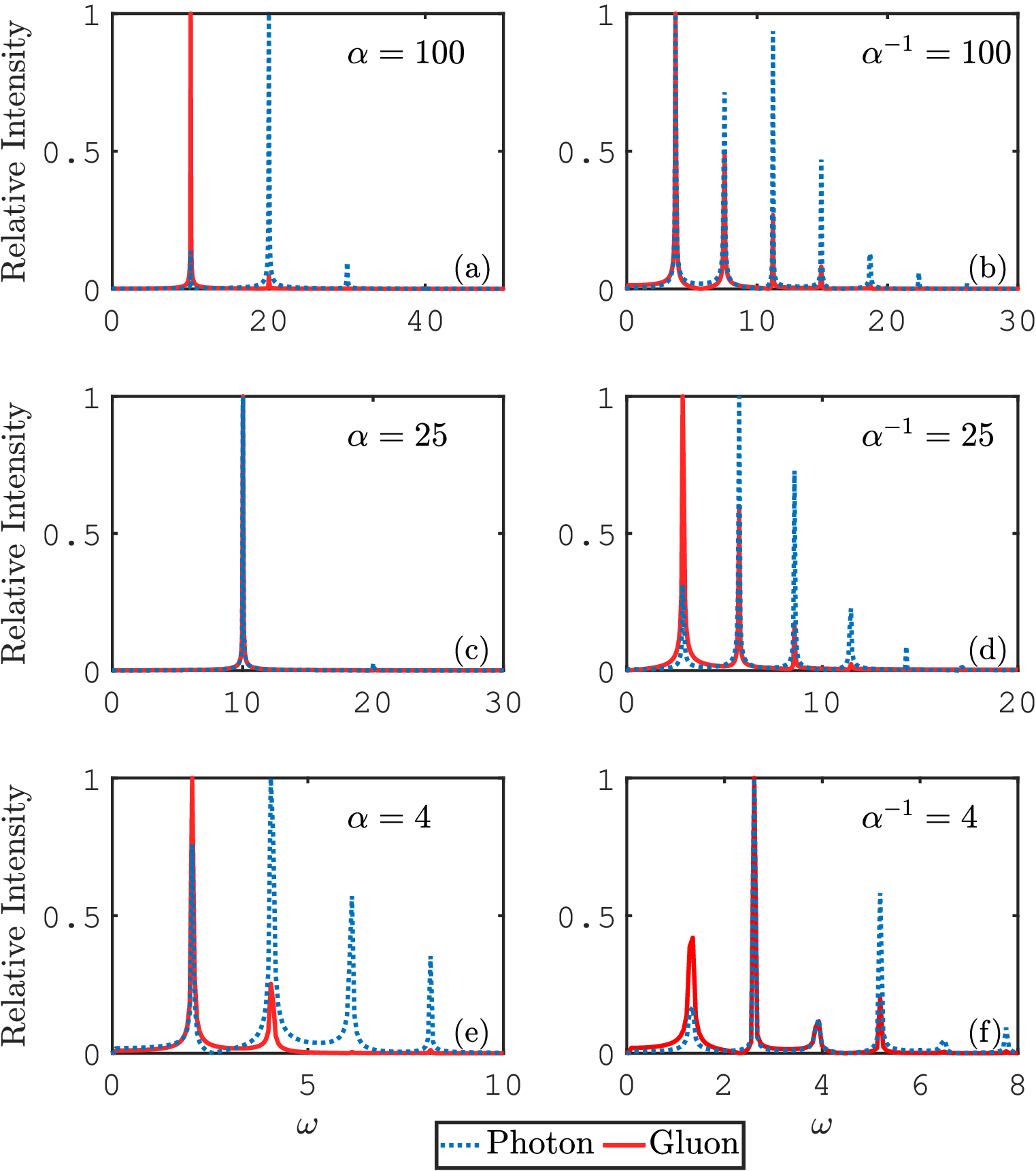}
    \caption{\justifying{Plots of energy spectra of gluonic(red) and photonic emissions (blue) in the same order as in Fig \ref{fig:1}, labelled respectively from $(a) - (f)$. }}
    \label{fig:2}
\end{figure}

\begin{figure}[h!]
    \centering
    \includegraphics[width=0.51\linewidth]{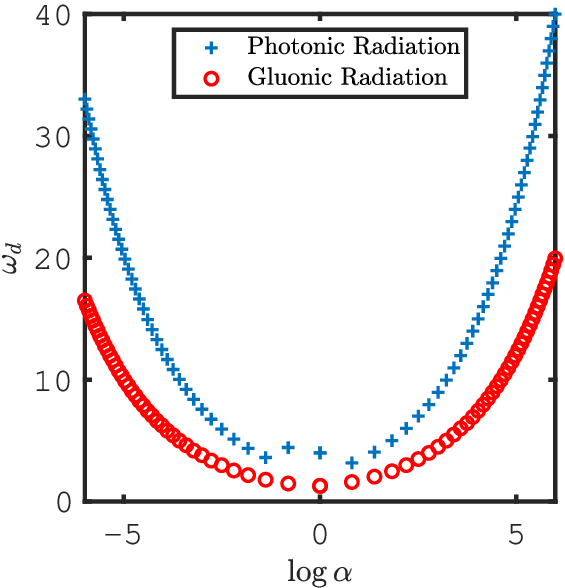}
    \caption{\justifying{The dominant frequencies    for  gluonic and  photonic radiations are  shown as  functions of $\alpha$ under the same  kinematic conditions}}
    \label{fig:3}
\end{figure}

A number of far-reaching conclusions can be drawn directly from (\ref{eq:consts}).  The foremost of them is that the particle kinetic energy and momenta are always bounded. The bounds become increasingly stringent in the two extreme limits $\alpha \rightarrow 0, \infty$.  Further,  $p_{min} \neq -p_{max}$, unless $\Pi = 0$. Thus, the mean position of the particle averaged over a period, drifts almost linearly away from its initial position, except when $\Pi = 0$ which leads to oscillatory motion. Finally, depending on the sign of $\Pi$, the particle moves either parallel to or opposite to the direction of the electric field. All these counterintuitive results stand in sharp contrast with electrodynamics where the particle is obliged to undergo a constant change in its momentum.

Moving on to details,  it follows from  Eq (\ref{effdyn}) that  there is  yet another constant of motion given by
\begin{eqnarray}
    \label{eq:const3}
    {\cal C}  &= & \frac{1}{2}\left( \frac{dp}{d\tau} \right)^2 + V_{eff}(p);  \nonumber \\
    V_{eff}(p) & = & \frac{1}{2} k_1 p^2  - k_2 \left(1 + p^2 \right)^{1/2} - k_3 p
\end{eqnarray}
The effective potential is absolutely confining for  $p$, as required. Equally pertinently, the shape of $V_{eff}(p)$ is dependent on the choice of the gauge potentials — the control parameter $\alpha$, impacting the particle dynamics strongly.  
   The results of numerical
calculations (with $\alpha$ =1 corresponding to the choice $\varphi = a =1$) performed in three regimes: $\alpha^2 \gg  1,  \alpha^2 \ll 1, \alpha^2 \simeq 1$,  for  $\Pi \gtrless 0, \Pi=0$  are illustrated in Fig \ref{fig:1},  which shows the phase space plots of the particle motion corresponding to six representative effective potentials.  They are so arranged (from top to bottom,  two each) to show drifts along the field direction ($\Pi > 0$), bounded orbits ($\Pi =0$), and drifts opposite to the direction of the field ($\Pi < 0$). The numerical results also confirm (not shown here) the qualitative observation that in the top two cases ($\vert \alpha \vert\gg 1$), the drift velocity is almost a constant. None of the phase space plots conform to the uniform rate of change of momentum mandated in the electrodynamic case.

Rich that the dynamics is, it is well nigh impossible to observe the trajectory directly, be it with quarks or in spintronics. The best way out is to employ the bremsstrahlung radiation emitted by the test particle for diagnostics. Since the particles also carry electric charge, we work out both gluonic and electromagnetic bremsstrahlung radiations. The latter is naturally more favoured since gluons are themselves not directly observable. 
For the sake of simplicity, we shall keep only the linear terms in the Lienard–Wiechert potential. We denote by $\dot{u}, \theta, \xi$ the acceleration,  and the angles made by the radiation with momentum and acceleration respectively.  Consider the gluon emission first. Paying due attention to the dynamical nature of the color charge, the angular distribution of the power of the emitted gluonic radiation may be expressed as a sum of two terms
\begin{equation}
\frac{d \mathcal{P}}{d \Omega} = \frac{d \mathcal{P}_c}{d \Omega} + \frac{d \mathcal{P}_y}{d \Omega}
\end{equation}
where the first term is common to photon emission and is given by the familiar expression (with $\vert \zeta\vert^2 \equiv 1$)
\begin{equation}
\frac{d \mathcal{P}_c}{d \Omega} = \frac{1}{4\pi} \frac{\vert\Vec{\zeta}\vert^2 \dot{u}^2}{\left(1 - u \cos\theta \right)^6} \Big( u \cos\xi \sin\theta + (1 - u \cos\theta) \sin\xi  \Big)^2 
\end{equation}
The second term, which is characteristic of  gluonic emission, incorporates the dynamical nature of the radiating color charge and is
given by 
\begin{equation}
\label{eq:rad}  
  \frac{d \mathcal{P}_y}{d \Omega} =  \frac{1}{4\pi}\frac{\vert\dot{\Vec{\zeta}}\vert^2 u^2 \sin^2\theta }{\left(1 - u \cos\theta \right)^4} 
    \end{equation}
   with     
\begin{equation}
    \label{eq:zdot}
    \vert\dot{\vec{\zeta}}\vert^2 = \dot{p}^2 \left(\left\{ \frac{(\Pi - p)}{\zeta_1 a^2} + \frac{p (\varepsilon - \gamma)}{\zeta_1 \gamma \varphi^2} \right\}^2 + \frac{1}{a^2} + \frac{p^2}{\gamma^2 \varphi^2} \right)
\end{equation}
Leaving aside the details of angular distribution, we look at the total power radiated which may be seen to be
\begin{equation}
    \label{eq:totpow}
    \mathcal{P} = \frac{2}{3} \vert \vec{\zeta} \vert^2 \dot{u}^2 \gamma^6 + 2  |\dot{\Vec{\zeta}}|^2 \gamma^2 - \frac{|\dot{\Vec{\zeta}}|^2}{u} \ln{\left({\frac{1 + u}{1-u}}\right)}   
\end{equation}
The corresponding expression for the electromagnetic bremsstrahlung  is obtained  by keeping only the first term  and setting 
 $\vert\vec{\zeta}\vert^2  \equiv e^2$,  which simply leaves us with the Larmor formula
$\mathcal{P}_{ED} = \frac{2e^2}{3}\dot{u}^2 \gamma^6$. 
The power spectrum is of particular interest.  Since both momentum and acceleration are strictly periodic, the spectrum is necessarily discrete and may be expected to carry decisive information on the underlying source.  Fig (\ref{fig:2}) shows the power spectra for both EM and gluonic radiations. The intensities for both of them are normalised relative to their respective maximum intensities.   One common feature that stands out is the richness of the spectral distribution. The spectral pattern in each case is distinct and acts as a unique signature of the underlying parameters, especially of $\alpha$. In fact, the most dominant frequency {$\omega_d$}, the frequency at which the relative intensities are of unit magnitude,  already conveys a wealth of information. This may be seen from Fig.\ref{fig:3} that, be it gluonic radiation or electromagnetic, $\omega_d$ resolves  $\alpha$ almost completely, except for a twofold discrete ambiguity, which may be resolved with a few additional measurements. Unsurprisingly, the power spectrum is asymmetric under $\alpha \rightarrow \alpha^{-1}$. 

In conclusion, we have taken up a seemingly textbook-like problem, and shown that charged particle dynamics holds great potential to shed light upon the non-abelian interactions, acting as a theoretical laboratory to explore nonintuitive features of YM dynamics even in the classical regime. The particle momentum in such fields is completely bounded and is deeply sensitive to the choice of the gauge potentials. This richness is reflected amply in the radiation (especially electromagnetic) emitted by the test particle. The power spectrum can act as a signature for getting information on the underlying sources. As averred, with further adaptation, one would be able to get better insight into YM dynamics ranging from spintronics, laser-matter interaction to QGP. 

\noindent\textbf{Acknowledgments:} The work of SB is supported by the Department of Science and Technology (DST), Govt. of India under DST/INSPIRE Fellowship/2019/IF190723. AD acknowledges support from the Science and Engineering Board (SERB) core grants CRG 2018/000624 and CRG/2022/002782 as well as J C Bose Fellowship grant JCB/2017/000055, Anusandhan National Research Foundation (ANRF), Government of India.

%
\end{document}